\documentclass[12pt]{iopart}

\usepackage{graphicx}
\usepackage{multirow}
\begin{document}

\title[Stable Charged Gravastar Model]{Stable Charged Gravastar model in Cylindrically Symmetric Space-time }

\author{D Bhattacharjee$^1$ and P K Chattopadhyay$^2$}

\address{$^{1,}$$^{2}$IUCAA Centre for Astronomy Research and Development (ICARD), Department of Physics, Coochbehar Panchanan Barma University, Vivekananda Street, District: Coochbehar, Pin: 736101, West Bengal, India}
\eads{\mailto{debadriwork@gmail.com}, \mailto{pkc$_{-}$76@rediffmail.com}}
\vspace{10pt}

\begin{abstract}
In this paper, we have extended the idea of gravitational Bose-Einstein condensate star (gravastar) to charged gravastar system and explored the role of charge in gravastar formation and its properties. We have used the most general line element in cylindrically symmetric space-time. In this approach the existence of singularity at the center of gravastar is removed and the event horizon is replaced by the thin shell approximation. The proper length of the shell is calculated along with the energy of the thin shell. A mass limit for thin shell has also been evaluated. The entropy calculation shows that the entropy of the configuration is smaller than that of a quasi-black hole system and even smaller than that of a classical black hole. Unlike black hole, the gravastar system is a stable configuration and there is no information paradox. 
\end{abstract}

%
\vspace{2pc}
\maketitle
%
%
\section{Introduction}
\label{intro}
In the last few decades there has been a wide research interest in understanding the issues both in Cosmology and Astrophysics. And for this reason, comapact objects are an important source as it provides a stage to test different relevant theories in high density regime. One of such compact objects are Gravastars. Recently, gravastars have acquired some attention mainly because of the fact that there is a connection between cosmology and the accelerating universe from the observational aspect. 
 
The study of astrophysical objects depend on the space-time metric. In general the spherically symmetric metric is used in case of charged as well as uncharged objeccts to study their internal physical properties. In this paper, we have used the cylindrically symmetric metric to construct the Einstein-Maxwell field equations for compact astrophysical objects. Safko and Witten \cite{Safko} in their solutions provided an overall comprihensive structure of the Einstein-Maxwell field equations coupled with electromagnetic field. Bronnikov et.al. \cite{Bronnikov}, provided the information regarding the Einstein field equations in the most general cylindrically symmetric metric. In another paper, Bronnikov \cite{Bronnikov1} coupled the solutions with electromagnetic field and calculated the Einstein-Maxwell field equations. 

The first model of Gravastar was proposed by Mazur and Mottola \cite{Mazur}. Later in another study, Visser and Wiltshire \cite{Visser} reduced the five layer model to a three layer model. In the final model, A gravastar has three layers. The interior region has the De-Sitter like space-time with an equation of state $p=-\rho$ while the exterior region is the vacuum space-time with $p=\rho=0$. In the interior region, gravity undergoes a vacuum rearrangement phase. The two space-times are separated by a thin shell approximation. The shell is characteried by the equation of state $p=\rho$. The thin shell replaces the notion of event horizon which is classical black hole feature. Thus, the gravastar removes all the inconsistencies imposed by a black hole.

With the advent of gravastar model, various conditions have been applied on it to check the formation and behaviour of such models. In a study, Yusaf \cite{Yusaf} constructed a cylindrical charged gravastar like structure and used modified gravity theories to examine the behaviour. In a similar study, Bhatti \cite{Bhatti} showed the stability of gravastars with cylindrical metric under linear perturbations around the equilibrium condition and different values of charge. Brandt \cite{Brandt} in the paper, constructed a model of charged gravastar with de-Sitter-Reissner-Nordstrom space-time and found that the existence of charge is important for the stability of the configuration and also they have showed that above the cetrain range of charge, the gravastar turns into a black hole. In another study, Ghosh et.al. \cite{Ghosh} have configured gravastars in higher dimension theories and found that application of higher dimensions in case of charged gravastars are not suitable on physical grounds. Yusaf et.al. \cite{Yusaf1} studied the overall charged gravastar model in spherically symmetric space-time, in modified gravity theory and they also examined the stability conditions. In this paper, we study the effect of radial electric field in case of charged gravastar in a most general cylindrically symmetric space-time. The solutions for the three regions are explored in presence of charge and the solutions have been used to model the gravastar extensively. All the physical parameters of a strong gravastar configuration have been studied in presence of charge and we have noted some new results. 

The paper is organised in the following manner. In \sref{intro}, we have given short introduction of charged gravastar. We have established the necessary mathematical formalism in \sref{mathform}. Under this section, \sref{IR}, \sref{TS} and \sref{ER} are devoted to calculate the $g_{rr}$ and $g_{tt}$ components for the three regions with respective equation of states. In \sref{BC}, we have matched the interior and thin shell solutions at $r=r_{1}$ and the thin shell and exterior region at $r=r_{2}$ and evaluated the values of different constants. \Sref{JC} deals with the junction conditions where the interior and exterior regions are matched over a smooth hypersurface. Through the matching condition, we find the mass of the thin shell as well. In \sref{gp}, we have shown the characteristic properties of a charged gravastar and their physical behaviour depending on the value of charge. In this section we have also calculated the proper length, thin shell energy, entropy of the stiff fluid and the variation of the equation of state of the thin shell with shell radius and charge. Finally, in \sref{D}, we discuss the main findings of the paper.  
     
\section{Mathematical Formalism}
\label{mathform} 
We use the static, cylindrically symmetric line element of the form  \cite{Bronnikov} given below:
\begin{center}
	\begin{equation}
		ds^2=e^{2\gamma} dt^2-e^{2\alpha} dr^2-e^{2\mu} dz^2-e^{2\beta} d\phi^2,\label{eq1}
	\end{equation}
\end{center}
here, $\gamma, \alpha, \mu, \beta$ are functions of $r$.
We have also considered a harmonic radial co-ordinate where $\alpha$ is represented in terms of the other three functions $\gamma, \mu, \beta$ as follows:
\begin{equation}
		\alpha=\gamma+\beta+\mu,\label{eq2}
\end{equation}
The Energy-Momentum tensor for matter sector is given as:
\begin{center}
	\begin{equation}
		({T}^{\mu}_{\nu})_{m}=diag (\rho,-p,-p,-p).\label{eq3}
	\end{equation}
\end{center}
Here, we have considered static charge distribution which is solely responsible for the gravitational field obeying the Maxwell's theory. The field is characterised by the antisymmetric tensor, $F_{\mu\nu}=-F_{\nu\mu}$. The Maxwell's equation is:
\begin{equation}
	\nabla_\eta F^{\eta\zeta}=4\pi j^{\zeta}, \label{eq4}
\end{equation}
where $j^{\zeta}$ is the eletric four-current and in vacuum $j^{\zeta}=0$. Incorporating the charge, the stress-energy tensor takes the form \cite{Bronnikov},  
\begin{equation}
	({T}^{\mu}_{\nu})_{q}=\frac{1}{4\pi} (-4F_{\mu\alpha}F^{\nu\alpha}+\delta_{\mu}^{\nu}F_{\alpha\beta}F^{\alpha\beta}). \label{eq5}
\end{equation}
Due to the emergence of the off-diagonal elements in $({T}^{\mu}_{\nu})_{q}$, \eref{eq4} restricts the simultaneous existence of the electric and magnetic fields in different directions at the same time allowing the similar directions for the fields. From \eref{eq3}, we get, 
\begin{equation}
	F^{\eta\zeta}= \frac{q_{\zeta}}{\sqrt{-g}}, \label{eq6}
\end{equation}
where, $q_{\zeta}$ is and constant and it is formulated as the total charge for the static charge distribution. From \eref{eq2} consideration, we can write, 
\begin{equation}
	\sqrt{-g}=e^{2\alpha}. \label{eq7}
\end{equation} 
Using \eref{eq6} and \eref{eq7} in \eref{eq5}, we get, 
\begin{equation}
	({T}^{\mu}_{\nu})_{q}=\frac{q^2}{2\pi} e^{2\gamma-2\alpha}. \label{eq8} 
\end{equation}
With the signature of the tensor as, 
\begin{equation}
	({T}^{\mu}_{\nu})_{q}=\frac{q^2}{2\pi} e^{2\gamma-2\alpha}diag(+,+,-,-) \label{eq9}
\end{equation}
Considering the static charge configuration, the Einstein's field equation (henceforth EFE) in presence of electric charge is written as:
\begin{equation}
		{R}^{\mu}_{\nu}-\frac{1}{2}{g}^{\mu}_{\nu}=-8\pi [({T}^{\mu}_{\nu})_{m}+({T}^{\mu}_{\nu})_{q}]. \label{eq10}
\end{equation}
From \eref{eq1}, we formulate the EFE as:
\begin{equation}
	\beta''+\mu''-\mu'\beta'-\mu'\gamma'-\beta'\gamma'=8\pi e^{2\alpha}[-\rho-\frac{q^2}{2\pi} e^{2\gamma-2\alpha}], \label{eq11}
\end{equation}
\begin{equation}
	\mu'\beta'+\mu'\gamma'+\beta'\gamma'=8\pi e^{2\alpha}[p-\frac{q^2}{2\pi} e^{2\gamma-2\alpha}], \label{eq12}
\end{equation}
\begin{equation}
	\mu''+\gamma''-\mu'\beta'-\mu'\gamma'-\beta'\gamma'=8\pi e^{2\alpha}[p+\frac{q^2}{2\pi} e^{2\gamma-2\alpha}], \label{eq13}
\end{equation}
\begin{equation}
	\gamma''+\beta''-\mu'\beta'-\mu'\gamma'-\beta'\gamma'=8\pi e^{2\alpha}[p+\frac{q^2}{2\pi} e^{2\gamma-2\alpha}]. \label{eq14}
\end{equation}
where $(')$ denotes the differentiation with respect to $r$. 
From a linear combination of \eref{eq11} to \eref{eq14}, we get a series of equations,
\begin{equation}
	\beta''+\gamma''=16\pi pe^{2\alpha}, \label{eq15}
\end{equation} 
\begin{equation}
2(\mu'\beta'+\mu'\gamma'+\beta'\gamma')-(\beta''+\gamma'')=-8q^2e^{2\gamma}, \label{eq16}
\end{equation}
\begin{equation}
	3\beta''+\gamma''+2\mu''-4(\mu'\beta'+\mu'\gamma'+\beta'\gamma')-(\beta''+\gamma'')=-16\pi\rho e^{2\alpha}. \label{eq17}
\end{equation}
From \eref{eq13} and \eref{eq14} we get, 
\begin{eqnarray}
	\beta''=\mu'', \nonumber \\ 
	\mu=\beta+ar, \label{eq18}	
\end{eqnarray}
where, a is an integration constant. \\
Let us consider, $\eta=\beta+\gamma$. Using this transformation the term $(\mu'\beta'+\mu'\gamma'+\beta'\gamma')$ can be rewritten as $(2\eta'\beta'+a\eta'-\beta'^{2})$, i.e.
\begin{equation}
	\mu'\beta'+\mu'\gamma'+\beta'\gamma'=2\eta'\beta'+a\eta'-\beta'^{2}. 
\end{equation}
To compute a general solution, we consider, $\rho=np$ \cite{Bronnikov} and a linear combination of \eref{eq11} to \eref{eq14} yields, 
\begin{equation}
	\frac{(n+1)}{4}\eta''-(2\beta'+a)\eta'+\beta''+\beta'^2=0. \label{eq20}
\end{equation}
Conservation of Stress-Energy Tensor, 
\begin{equation}
	\nabla_{l}T_{l}^{k}=0, \label{eq21}
\end{equation}
produces, 
\begin{equation}
	p'+(\rho+p)\gamma'=0. \label{eq22}
\end{equation}
Following the work \cite{Bronnikov1}, we have considered an arbitrary form of $\beta$ as , 
\begin{equation}
	\beta=Ar^2. \label{eq23}
\end{equation}
Gravastar is a three layered configuration that is characterised by three distinct equation of sates (EOS) given below-
\begin{itemize}
	\item Interior (Region-I): \hspace{1cm} $0\le r < r_1$ ,\hspace{0.5cm} $\rho=-p$
	
	\item Shell (Region-II): \hspace{1cm}$r_1< r<r_2$,\hspace{0.8cm} $\rho=p$ 
	
	\item Exterior (Region-III): \hspace{1cm}$r_2< r$,\hspace{0.9cm} $\rho=p=0$ 
\end{itemize}
\subsection{\bf Interior Region} \label{IR} Applying the equation of state for the interior region, $n=-1$ $(\rho=-p)$ in \eref{eq20}, we get, 
\begin{equation}
		\gamma''=8q^2e^{2\gamma}+2A, \label{eq24}
\end{equation}
Therefore, 
\begin{equation}
	\gamma=-\frac{A+4q^2}{8q^2}+c_1e^{4qr}+c_2e^{-4qr}. \label{eq25}
\end{equation}
Where, $c_1$ and $c_2$ are integration constants.Using \eref{eq18}, \eref{eq23} and \eref{eq25} in \eref{eq2}, we get, 
\begin{equation}
	\alpha=2Ar^2+ar-\frac{A+4q^2}{8q^2}+c_1e^{4qr}+c_2e^{-4qr}. \label{eq26}
\end{equation}
\Eref{eq25} and \eref{eq26} clearly shows that at the center ($r=0$) $\gamma$ and $\alpha$ are constants, which in turn proves that $g_{rr}$ and $g_{tt}$ pick up finite values, thereby removing the notion of singularity. 
\subsection{\bf Thin Shell}\label{TS} Thin shell region contains ultra-relativistic stiff fluid with the equation is state $\rho=p$. Therefore, we use $n=1$ in \eref{eq20}, 
\begin{equation}
	\frac{\eta''}{2}+\beta''=(2\beta'\eta'+a\eta'-\beta'^2), \label{eq27}
\end{equation}	
Using \eref{eq16} in \eref{eq27}, we finally get, 
\begin{equation}
	\gamma=\frac{\ln(-A/2q^2)}{2}. \label{eq28}
\end{equation}
Using \eref{eq18}, \eref{eq23} and \eref{eq28} in \eref{eq2}, we get,
\begin{equation}
	\alpha=r(a+2Ar)+\frac{\ln(-A/2q^2)}{2}. \label{eq29}
\end{equation}
Using \eref{eq28} in \eref{eq22} we get, 
\begin{equation}
	p=\rho=\frac{-2q^2}{A}. \label{eq30}
\end{equation}
We can see that, in the thin shell approximation $(r\rightarrow0)$, $g_{rr}$ and $g_{tt}$ approaches a non-vanishing constant value. Therefore, we can surely conclude that the thin shell replaces the event horizon in case of a charged gravastar. This feature differentiates a gravastar from a black hole. 
\subsection{\bf Exterior Region}\label{ER} The exterior region is the vacuum space-time with the equation of state, $\rho=p=0$. From \eref{eq15}, we get,  
\begin{eqnarray}
	\gamma''=-\beta''  \nonumber \\
	\Rightarrow \gamma=-Ar^2+br. \label{eq31}
\end{eqnarray}
Where, $b$ is an integration constant. Again using \eref{eq18}, \eref{eq23} and \eref{eq31} in \eref{eq2},
\begin{equation}
	\alpha=Ar^2+(a+b)r. \label{eq32}
\end{equation}
We compute the Kretschmann scalar $K_{S}$ to show that the exterior region is a flat space-time in a comprihensive way. The Kretschmann scalar $K_{S}$ is computed as, 
\begin{equation}
	K_{S}=R^{abcd}R_{abcd}. \label{eq33}
\end{equation} 
Where, $R$ is the Reimann Curvature Tensor. Kretschmann scalar for the exterior region is given as,
\begin{eqnarray}
	R_{S}=\frac{1}{4r^6(a+b+Ar)^4}(\frac{4(b-2Ar)^2(a+b+Ar)^2}{(b-Ar)^2}+ \nonumber \\
	\frac{4(a+2Ar)^2(a+b+Ar)^2}{(a+Ar)^2}\nonumber\\+\frac{(b-2Ar)^2(a+b+Ar)^2(a+2Ar)^2}{(b-Ar)^2(a+Ar)^2}+4(a+b+2Ar)^2+\nonumber\\ \frac{(2ab^2+2b^3-3aAbr+2aA^2r-4A^2br^2+4A^3r^3)^2}{(b-Ar)^4}+\nonumber\\
	\frac{(2a^3+2A^2r^2(b+2Ar)+2a^2(b+3Ar)+aAr(3b+8Ar))^2}{(a+Ar)^4} \label{eq34}
\end{eqnarray}
At $r\rightarrow\infty$, $R_{S}\rightarrow0$, proving that the exterior region is a flat vacuum space-time.
\section{Boundary Condition} \label{BC} Boundary conditions are a valid and verifiable method to explore the values of different constants used in the context. Here, we equate the interior and thin shell region at the boundary $r=r_{1}$ and evaluate the constant $A$ and also find the dependence of $A$ on charge $q$. Similarly, we match the thin shell and the exterior region at $r=r_{2}$, to find the range of the outer radius hence the range of thickness $(r_{2}-r_{1})$ of charged gravastar. We take the interior radius $R_{1}=10 km$ \cite{BCP}. 
\begin{itemize}
	\item Matching interior and thin shell at $r=r_{1}=10~km$: 
	\begin{equation}
	-\frac{A+4q^2}{8q^2}+c_1e^{4qr}+c_2e^{-4qr}=\frac{\ln(-A/2q^2)}{2}, \label{eq35}
	\end{equation}
	\begin{equation}
\fl2Ar_1^2+ar_1-\frac{A+4q^2}{8q^2}+c_1e^{4qr_1}+c_2e^{-4qr_1}=r_{1}(a+2Ar_{1})+\frac{\ln(-A/2q^2)}{2}. \label{eq36}
	\end{equation} 
\item Matching the thin shell and exterior region at $r=r_{2}$: 
\begin{equation}
\frac{\ln(-A/2q^2)}{2}=-Ar_{2}^2+br_{2}, \label{eq37}
\end{equation} 
\begin{equation}
	r_{2}(a+2Ar_{2})+\frac{\ln(-A/2q^2)}{2}=Ar_{2}^2+(a+b)r_{2}, \label{eq38}
\end{equation} 
\end{itemize}
\begin{table}
\caption{\label{t1} Shell thickness~$\epsilon~(Km)$ and dependence of constant A on charge q for $r_1=10~(Km)$ and $b=0.00006$.}
\vspace{0.5cm}
\begin{indented}
\item[]\begin{tabular}{@{}|c|c|c|c|}
\hline
Charge(q) & A & Charge(q) & A  \\ \hline
\multirow{5}{*}{0.01} 	& -0.000192657 & \multirow{5}{*}{0.05} & -0.00283584 \\  
						& -0.000192636 && -0.00283276 \\ 
						& -0.000192614 && -0.00282969 \\
						& -0.000192593 && -0.00282663  \\
						& -0.000192572 && -0.00282356 \\ \hline 
\multirow{5}{*}{0.03}	& -0.00136960  &\multirow{5}{*}{0.07} & -0.00421556\\ 
						& -0.00136872  && -0.00420979\\ 
						& -0.00136784  && -0.00420402  \\
						& -0.00136696  && -0.00419827 \\
						& -0.00136608  && -0.00419253\\ \hline    
\end{tabular}
\end{indented}
\end{table}
Here, we take the values of $a=0.00006$ \cite{Yusaf}, $r_{1}=10~km$\cite{BCP} and obtain the values of \lq$A$' numerically for different values of charge $q$. Also keeping in mind the range of the thickness as described in \cite{BCP}, we have used a range of thickness, $r_2$ from $10.01 km$ to $10.07 km$. The range is an arbitrary one that has been considered as per the thin shell approximation of a gravastar. Using \eref{eq37} and \eref{eq38}, we get the values of $A$ which we enlist in \tref{t1}. Here we have restricted our charge limit to $q=0.01 (km)$ to $q=0.07 (km)$. As we go higher than the limit, the mass of the shell becomes negative for $q=0.2 (km)$ and for values lower than $q=0.07 (km)$, the mass of the shell gets smaller. Therefore, we have considered this particular range of charge for this formulation.     
\section{Junction Condition} \label{JC} Gravastar has three distinct regions characterised by three equations of state as mentioned before. We match the interior region and the exterior region over a smooth hypersurface $\Omega$ at $r=D$. We have studied that the metric coefficients are continuous across the junction while calculating the boundary conditions. Whether, their derivatives are also continuous or not are subjected to investigation. Now we will use Darmois-Israel condition, \cite{DI}-\cite{Israel1} to compute the surface stresses at the junction interface. The intrinsic surface energy tensor ${S}_{ij}$ is given by Lanczos equation \cite{Lanczos}-\cite{Musgrave} of the form, 
\begin{equation}
	{S}^{i}_{j}=-\frac{1}{8\pi} ({K}^{i}_{j}-{\delta}^{i}_{j}{K}^{k}_{k}), \label{eq39}
\end{equation}
Here, ${K}_{ij}={K}^{+}_{ij}-{K}^{-}_{ij}$ where $(+)$ and $(-)$ sign indicate the exterior and interior interfaces respectively. The second fundamental form is given as-
\begin{equation}
	{K}_{ij}^{\pm}=-{\eta}_{ij}^{\pm} (\frac{\partial^2 X_{\nu}}{\partial \zeta^{i} \partial \zeta^{j}}+{\Gamma}_{\alpha\beta}^{\nu} \frac{\partial X_{\alpha}}{\partial \zeta^{i}} \frac{\partial X_{\beta}}{\partial \zeta^{j}} ). \label{eq40}
\end{equation}
The double sided normal on the surface is defined as, 
\begin{equation}
	\eta_{ij}^{\pm}={\pm} ({g}^{\alpha\beta} \frac{\partial f}{\partial x\alpha} \frac{\partial f}{\partial x\beta})^{1/2}, \label{eq41}
\end{equation}
with $\eta^{\nu}\eta_{\nu}=1$. Following Lanczos equation, the surface stress-energy tensor at the boundary of the interface is, ${S}_{ij}=diag(\varrho, -\vartheta, -\vartheta, -\vartheta )$, where $\varrho$ and $\vartheta$ are surface energy density and surface pressure respectively. The parameters $\varrho$ and $\vartheta$ are expressed as, 
\begin{eqnarray}
	\varrho=-\frac{1}{4\pi R}{\Big(\sqrt{f(r)}\Big)}^{+}_{-}~, \label{eq42} \\
	\vartheta=-\frac{\varrho}{2}+\frac{1}{16\pi} \Big(\frac{f'(r)}{\sqrt{f(r)}}\Big)^{+}_{-}. \label{eq43}
\end{eqnarray}
For the exterior region $f(r)=e^{2r(b-Ar)}$ and for the interior region $f(r)=e^{2(-\frac{A+4q^2}{8q^2}+c_1e^{4qr}+c_2e^{-4qr})}$. At $r=D$, \Eref{eq42} reduces to, 
\begin{equation}
\varrho=-\frac{1}{4\pi D}\Big(\sqrt{e^{2D(b-AD)}}-{\sqrt{e^{2(-\frac{A+4q^2}{8q^2}+c_1e^{4qD}+c_2e^{-4qD})}}}~\Big). \label{eq44}
\end{equation}
Hence, the mass of the thin shell is calculated as, 
\begin{equation}
 M_{shell} = 4\pi D^2\varrho= D\Big({\sqrt{e^{2(-\frac{A+4q^2}{8q^2}+c_1e^{4qD}+c_2e^{-4qD})}}}-\sqrt{e^{2D(b-AD)}}~\Big). \label{eq45}
\end{equation}
\section{Properties of charged gravastar} \label{gp}
\subsection{Proper length of the thin shell} Thin shell is the partition for the interior space and the exterior vacuum region. It is calculated as, 
\begin{eqnarray}
	l=\int_{r_{1}}^{r_{2}} e^{\alpha} dr \nonumber \\
	\Rightarrow l=\int_{r_{1}}^{r_{2}} e^{r(a+2Ar)+\frac{\ln(-A/4q^2)}{2}} dr. \label{eq46}
\end{eqnarray} 
Using \tref{t1}, we show the variation of length with increasing shell thickness. 
\begin{figure}[h]
	\centering
	\includegraphics[width=0.48\textwidth]{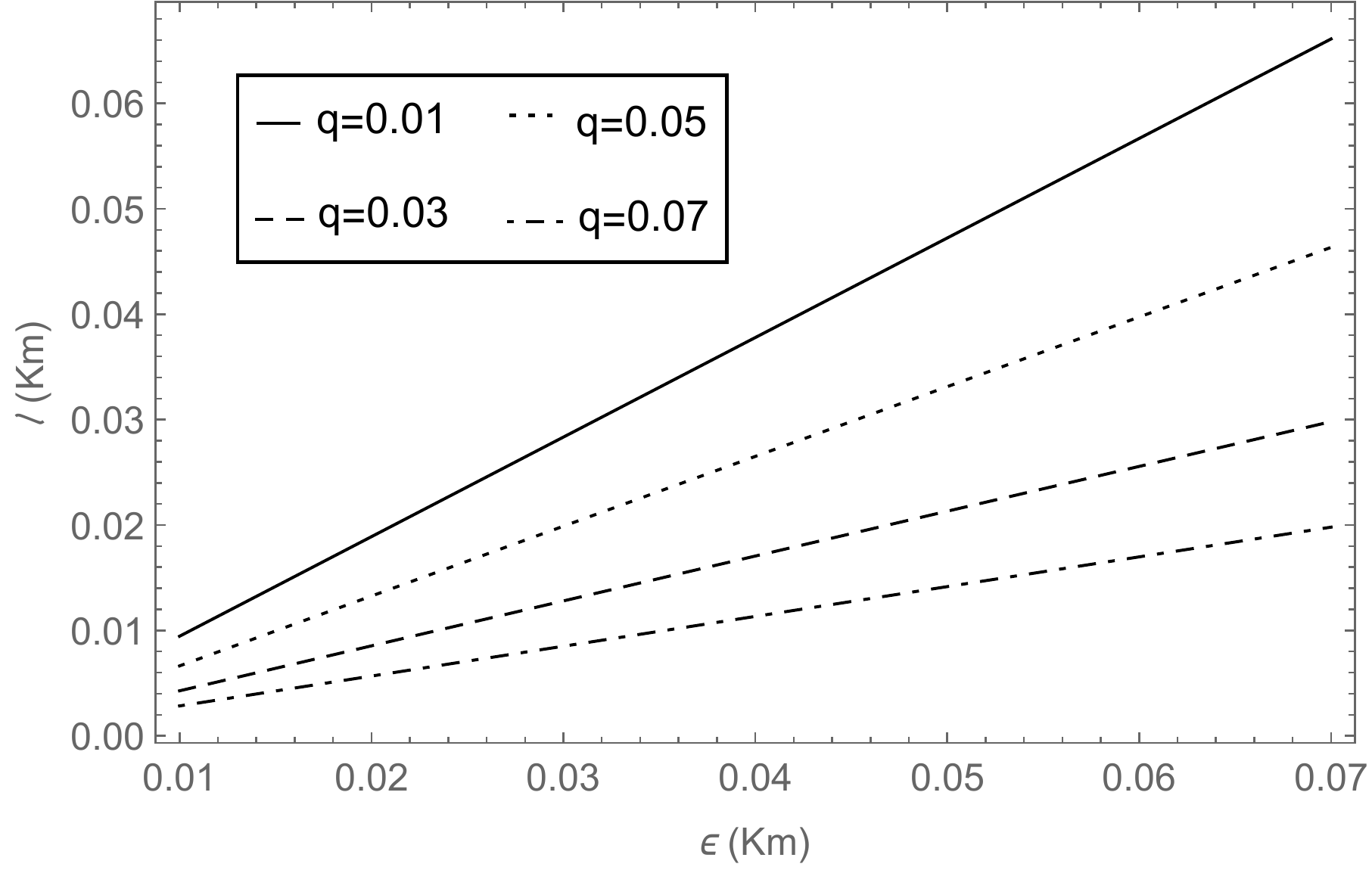}
	\caption{Variation of proper length of gravastar shell $l~(Km)$ with shell thickness $\epsilon~(Km)$.}
	\label{fig2}
\end{figure}
From \fref{fig2}, it is evident that the proper length of the gravastar shell decreases with the increase of charge $q$ i.e. as the charge in the shell is increased, the proper lenth decreases, which implies that less charged thin shell has a greater proper length. This feature is also described in \cite{Yusaf}.  
\subsection{Energy within the shell}The energy content withing the thin shell is given as \cite{Yusaf}, 
\begin{equation}
	E= \int_{r_1}^{r_2} 4\pi \rho r^2 dr, \label{eq47}
\end{equation}
using \eref{eq30} in \eref{eq47}, we get, 
\begin{eqnarray}
	E=-\frac{8\pi q^2}{A} \int_{r_1}^{r_2} r^2 dr, \nonumber \\
\Rightarrow	E=-\frac{8\pi q^2}{3A} [r^3]^{r_2}_{r_1}. \label{eq48}
\end{eqnarray}
\begin{figure}[h]
	\centering
	\includegraphics[width=0.48\textwidth]{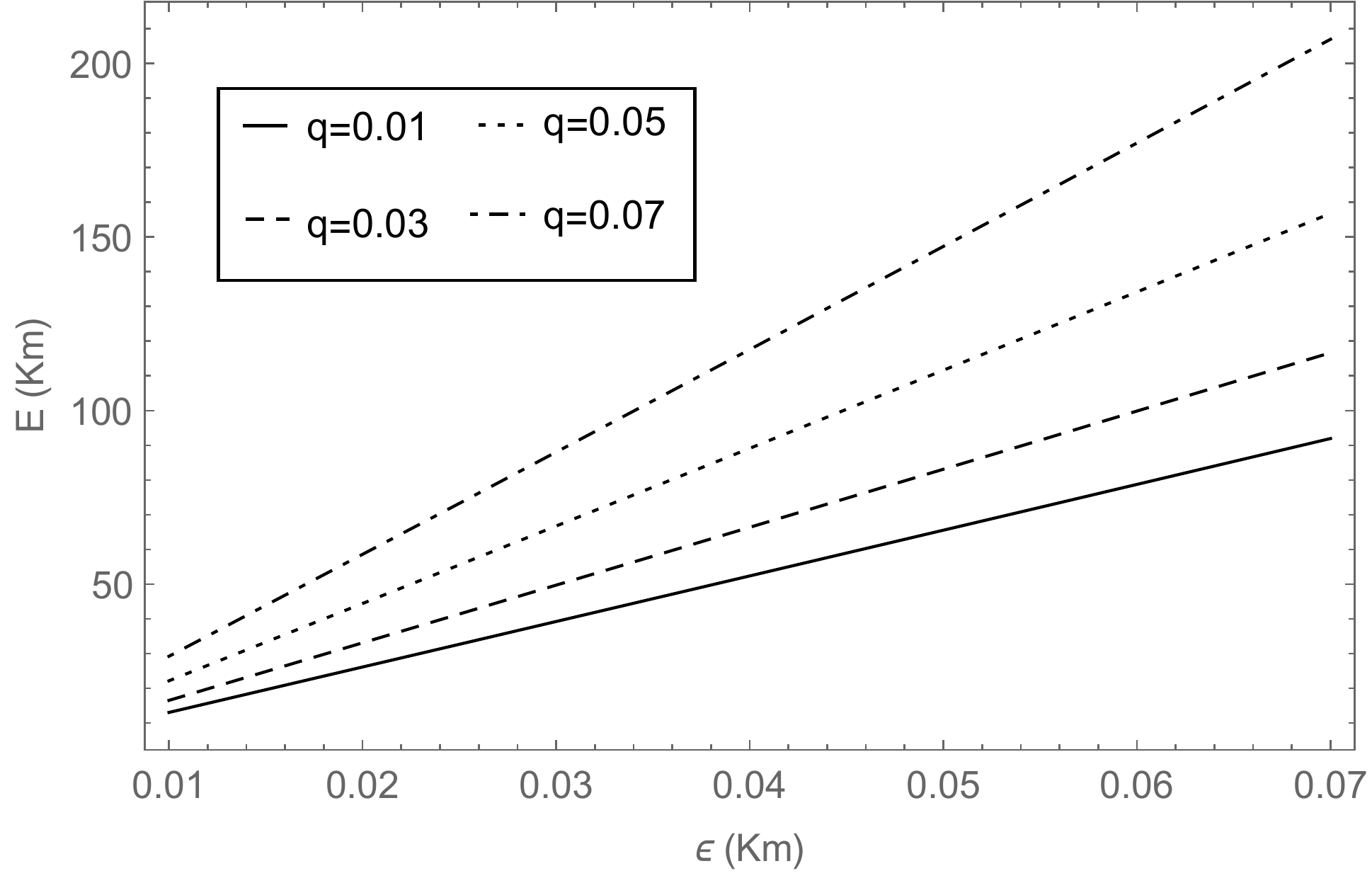}
	\caption{Variation of energy of gravastar shell $E~(Km)$ with shell thickness $\epsilon~(Km)$ and charge q(Km).}
	\label{fig3}
\end{figure}
With the use of \tref{t1}, \fref{fig3} describes the behavior of the energy contained within the thin shell with increasing thickness. It is evident that the energy increases as the quantity of charge in the shell is increased. 
\subsection{Entropy of the shell} Entropy is the manifestation of disorderness of the gravastar system. In literature it is evident that the entropy density of the gravastar interior is zero. The entropy of the thin shell is obtained by the following formula:
\begin{equation}
	S=4\pi\int_{r_1}^{r_2} s(r)r^2e^{\alpha} dr, \label{eq49}
\end{equation}
Where, $s(r)=\frac{\sigma k_B}{\hbar}\sqrt{\frac{p}{2\pi}}$ is the entropy density, and from \eref{eq30} we get, 
\begin{equation}
S = \frac{4\pi k_B \sigma}{\hbar\sqrt{2\pi}}\sqrt{\frac{-2q^2}{A}}\int_{r_1}^{r_2} r^2e^{\alpha}dr, \label{eq50}	
\end{equation}
using \eref{eq29} in \eref{eq50}, 
\begin{equation}
		S = \frac{4\pi k_B \sigma}{\hbar\sqrt{2\pi}}\sqrt{\frac{-2q^2}{A}}\int_{r_1}^{r_2} r^2  e^{r(a+2Ar)+\frac{\ln(-A/4q^2)}{2}} dr. \label{eq51}	
\end{equation}
Here, $\sigma$ is a dimenstionless constant. Again, we use the values enlisted in \tref{t1} to evaluate the behaviour of shell entropy with increasing thickness. 
\begin{figure}[h]
	\centering
	\includegraphics[width=0.48\textwidth]{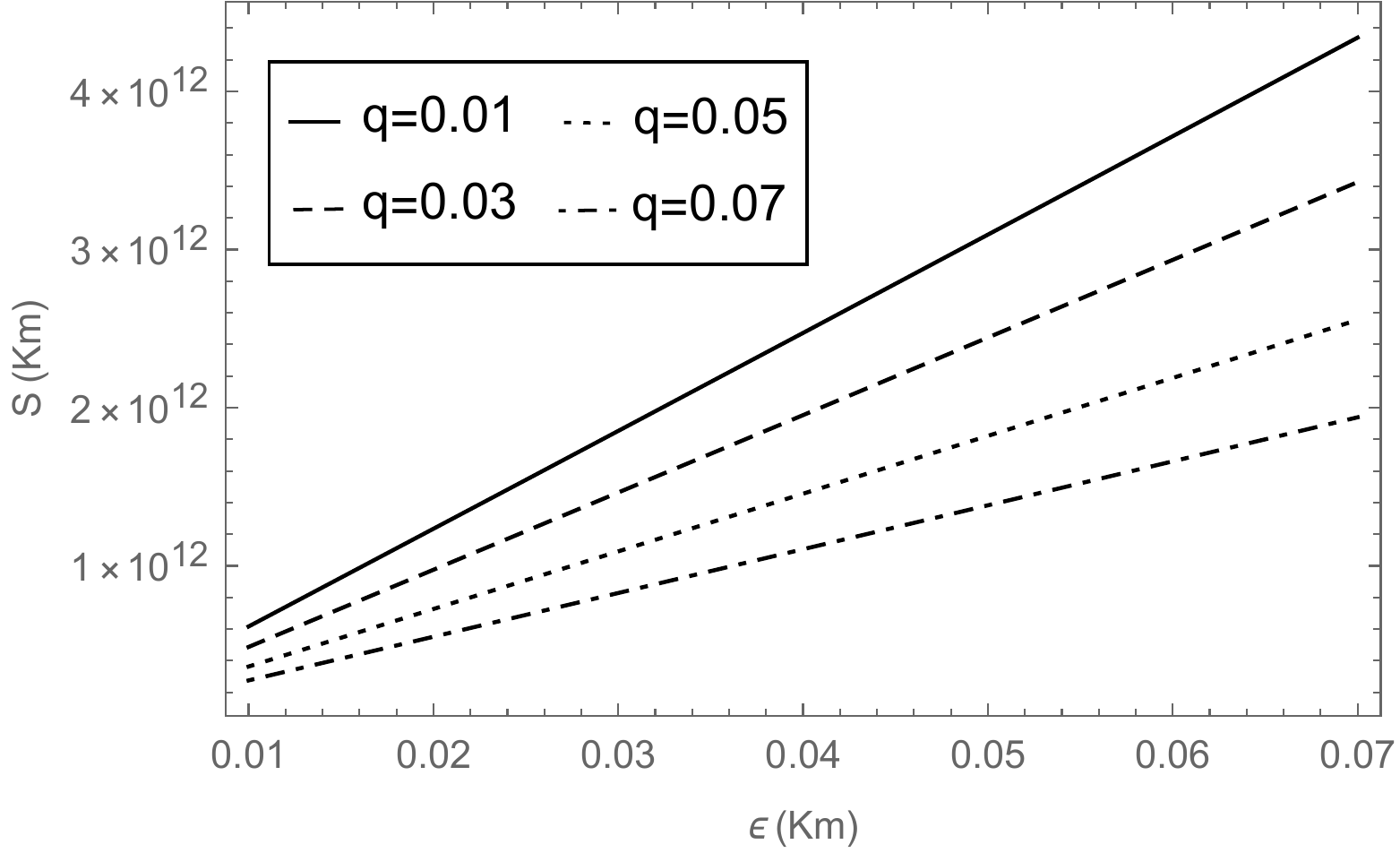}
	\caption{Variation of entropy of gravastar shell $S~(Km)$ with shell thickness $\epsilon~(Km)$ with $\sigma=1$ and charge q (Km).}
	\label{fig4}
\end{figure}
\Fref{fig4} shows that the more charged fluid contanis lower entropy \cite{Yusaf}. The entropy is even smaller than that of a quasi-black hole entropy as described in \cite{Mazur}.
\subsection{Equation of State} The equation of state for the thin shell is given by \eref{eq30} as, 
\begin{equation}
	p=\rho=\frac{-2q^2}{A}. \label{52}
\end{equation}
\begin{figure}[h]
	\centering
	\includegraphics[width=0.85\textwidth]{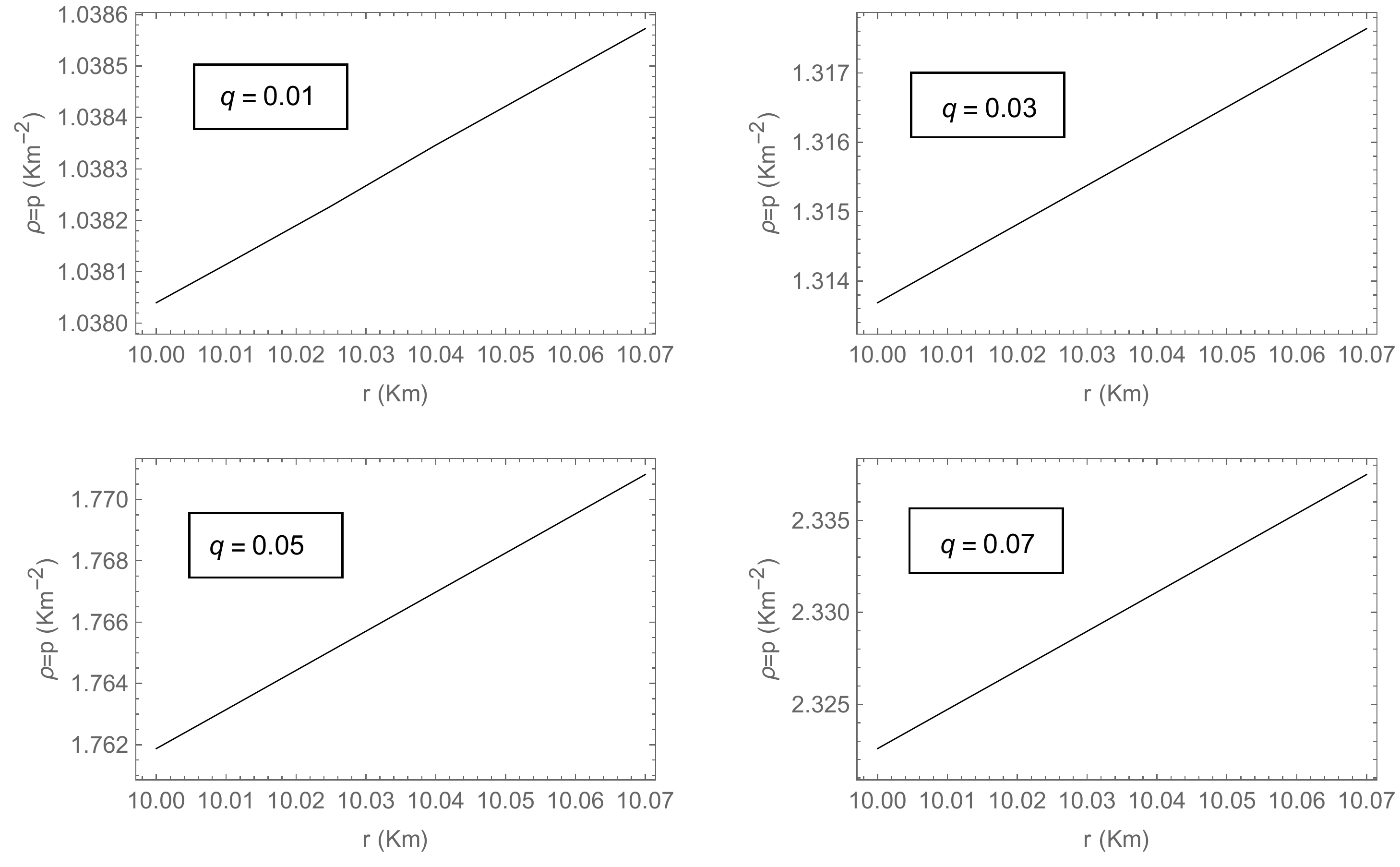}
	\caption{Variation of equation of state $p=\rho~(Km^2)$ with shell radius $r~(Km)$ and charge q (Km).}
	\label{fig5}
\end{figure}
\Fref{fig5} shows that the density of the ultra-relativistic fluid within the shell increases linearly with increasing radius. The density is also proportional to the charge \eref{52}. Therefore, towards the outer radius of the shell, the fluid is denser as well as highly charged \cite{Yusaf}.
\section{Discussion} \label{D} In this paper we have investigated the existence of gravitationally bound vacuum star or gravastar in the most general cylindrically symmetric spcae-time as described in \cite{Bronnikov} and also we have explored the role of electromagnetic field (radial electric field) in the formation of charged gravastar. The gravastar formalism is established by extending the concept of Bose-Einstein condensate to gravitational systems\cite{Mazur}. The gravastar is described by three distinct regions having three different equations of state. The first layer is the interior de-Sitter like space-time that removes any notion of singularity which is present in case of a classical black hole. The outer region is the vacuum space-time and the event horizon is replaced by the thin shell approximation, which separates the interior and exterior geometries. This is the basic configuration of a gravastar. We set up the Einstein-Maxwell's equations in the cylindrical form \cite{Bronnikov1} and solved it for the three regions. We have studied how the presence of charge affects the formation and behaviour of a gravastar. We have also pointed out the different properties, i.e. proper length, energy of the shell, entropy of the shell and the equation of state, that are pertinent in case of a charged gravastar. We found that, as the charge is increased, the proper length of the shell decreases, the energy of the shell increases and the entropy of the stiff fluid contained within the shell decreases. Moreover, the density of the ultra-relativistic fluid increases with increasing charge and shell radius which indicates that as we go towards the outer edge of the shell, denser and more charged fluid is found. The results are in accordance with \cite{Yusaf1}. We have also been able to show a mass limit for the thin shell using the Darmois-Israel junction condition \cite{DI}-\cite{Israel1}. Finally, we can assert that a gravastar discards all the inconsistencies of a physical theory that are imposed by a black hole and emerges as a strong alternative to black holes.

\end{document}